\documentclass[12pt]{article}
\pdfoutput=1
\usepackage{subfigure}
\usepackage{amssymb,amsmath}
\usepackage{graphicx}
\usepackage{color}
\usepackage[colorlinks=true
,urlcolor=blue
,citecolor=blue
,linkcolor=blue
,pagecolor=blue
,linktocpage=true
,pdfproducer=medialab
]{hyperref}
\usepackage[a4paper,width=15.2cm]{geometry}

\makeatletter \renewcommand{\@dotsep}{10000} \makeatother

\newcommand{\beq}{\begin{equation}}
\newcommand{\eeq}{\end{equation}}
\newcommand{\bea}{\begin{eqnarray}}
\newcommand{\eea}{\end{eqnarray}}

\begin{document}

\begin{center}

 {\Large\bf  Split Sfermion Families, Yukawa Unification and Muon $\boldsymbol {g-2}$
 } \vspace{1cm}

{\large   M. Adeel Ajaib\footnote{ E-mail: adeel@udel.edu}, Ilia Gogoladze\footnote{E-mail: ilia@bartol.udel.edu\\
\hspace*{0.5cm} On  leave of absence from: Andronikashvili Institute
of Physics, 0177 Tbilisi, Georgia.},  Qaisar Shafi\footnote{ E-mail:
shafi@bartol.udel.edu} and Cem Salih $\ddot{\rm U}$n \footnote{
E-mail: cemsalihun@bartol.udel.edu}} \vspace{.5cm}

{\baselineskip 20pt \it
Bartol Research Institute, Department of Physics and Astronomy, \\
University of Delaware, Newark, DE 19716, USA  } \vspace{.5cm}

\vspace{1.5cm}

 {\bf Abstract}
\end{center}

 We consider two distinct classes of Yukawa unified supersymmetric SO(10) models with non-universal and universal soft supersymmetry breaking (SSB) gaugino masses at $M_{\rm GUT}$. In both cases, we assume that the third family SSB sfermion masses at $M_{\rm GUT}$ are different from the corresponding sfermion masses of the first two families (which are equal). For the SO(10) model with essentially arbitrary (non-universal) gaugino masses at $M_{\rm GUT}$, it is shown that $t$-$b$-$\tau$ Yukawa coupling unification is compatible, among other things, with the 125 GeV Higgs boson mass, the WMAP relic dark matter density, and with the resolution of the apparent muon $g-2$ anomaly. The colored sparticles in this case all turn out to be quite heavy, of order 5 TeV or more, but the sleptons (smuon and stau) can be very light, of order 200 GeV or so. For the SO(10) model with universal gaugino masses and NUHM2 boundary conditions, the muon $g-2$ anomaly cannot be resolved. However, the gluino in this class of models is not too heavy, $\lesssim$ 3 TeV, and therefore may be found at the LHC.

\newpage

\renewcommand{\thefootnote}{\arabic{footnote}}
\setcounter{footnote}{0}



\section{\label{ch:introduction}Introduction}

 {Even though supersymmetric particles have not {yet} been observed, low scale supersymmetry (SUSY) remains at the forefront of beyond the Standard Model (SM) physics scenarios. In addition to {resolving} the gauge hierarchy problem and accommodating radiative electroweak symmetry breaking (REWSB), {SUSY} also provides a {compelling} dark matter candidate (the lightest supersymmetric particle (LSP)).  Contrary to the  non-supersymmetric case, {the three} gauge couplings naturally unify \cite{Dimopoulos:1981yj} around  $10^{16}$ GeV ($M_{\rm GUT}$), which therefore provides {an additional reason to suspect that SUSY may be found soon, hopefully at LHC 14}.}

 {It is well known that} gauge coupling unification does not significantly constrain the sparticle spectrum. On the other hand, {imposing} $t$-$b$-$\tau$ Yukawa coupling unification condition at $M_{\rm GUT}$ \cite{yukawaUn} can place significant constraints on the supersymmetric spectrum in order to fit the top, bottom and tau masses.
These constraints {are quite} severe \cite{Baer:2012cp,Gogoladze:2012ii,Gogoladze:2011aa,Ajaib:2013zha}, especially  after the discovery of a SM like Higgs boson with mass, $m_h \simeq 125 - 126$ GeV  \cite{:2012gk,:2012gu}.

{The constraints} from $t$-$b$-$\tau$ Yukawa coupling unification  depend
on the particular boundary conditions at $M_{\rm GUT}$ for the soft supersymmetry breaking (SSB) parameters \cite{susy-thres}.
{To be more precise, $t$-$b$-$\tau$ Yukawa unification is {successfully} realized if the threshold  corrections to the bottom quark mass are {suitably} large and have the correct sign.} The dominant contributions {arise} from {loop corrections involving the} gluino ($m_{\tilde g}$), the third generation sfermions and {the SSB} trilinear interactions \cite{susy-thres}.
 On the other hand, a 125 GeV  light CP-even Higgs boson mass also requires large radiative corrections, and the dominant contributions in this case also arise from third generation sfermions  and trilinear SSB scalar interaction \cite{at}.
Thus,  $t$-$b$-$\tau$ Yukawa unification and the 125 GeV light CP even Higgs boson {together} strongly constrain the {gluino and} third generation sfermion masses  {as well as the trilinear SSB couplings. (For a recent discussion regarding the top quark mass and related issues in low scale supersymmetric models, see ref. \cite{Gogoladze:2014hca})}

{We consider} two {choices} for the minimal set of SSB parameters  at  $M_{\rm GUT}$
which can lead to $t$-$b$-$\tau$ Yukawa unification. The first case {has} universal SSB gaugino mass terms but non-universal Higgs SSB terms, $m^2_{H_{u}}\neq m^2_{H_{d}}$ \cite{Blazek:2001sb}. Here $m_{H_{u},H_d}$  {denote the} up/down type Higgs SSB masses. In this case $t$-$b$-$\tau$  Yukawa unification can be realized {if the} gluino mass ($M_{\tilde g}$) is much smaller than the sbottom quark mass ($m_{\tilde {b}}$), and the stop trilinear SSB term ($A_t$) is larger than the stop mass ($m_{\tilde {t}}$).  To realize a 125 GeV light CP even Higgs boson in this scenario we {require} $M_{\tilde g} \leq 3$ TeV and $m_{\tilde {b}} \geq 10$ TeV. This also yields bounds on the  fundamental SSB parameters, {namely},  $m_0\gtrsim 10$ TeV and $M_{1/2}\lesssim 1$ TeV \cite{Baer:2012cp}.  Here $m_0$ and $M_{1/2}$  are   ${\rm GUT}$ scale universal SSB mass terms for the sfermions and  gauginos, respectively.

The second  class of {SO(10)} models have universal SSB Higgs  mass$^2$ term ($m^2_{H_{u}}= m^2_{H_{d}}$), whereas the gaugino SSB masses are non-universal
   at $M_{\rm GUT}$  \cite{Gogoladze:2009ug}. In this scenario the desired supersymmetric threshold  corrections
to $t$-$b$-$\tau$ Yukawa couplings can be realized {with} $M_{\tilde g} \gtrsim m_{\tilde {b}}$ \cite{Gogoladze:2012ii}.
For a particular choice of SSB gaugino masses ($M_1:M_2:M_3=1:3:-2$) at $M_{\rm GUT}$, which can be derived in
the framework of SO(10) GUT, it was {shown} \cite{Gogoladze:2011aa,Ajaib:2013zha}
that the  CP-even SM-like Higgs boson mass $m_h\approx 125$ GeV can be predicted from $t$-$b$-$\tau$ YU.  This result does not change much in terms of the Higgs mass prediction  {if} we relax $t$-$b$-$\tau$ YU up to $10\%$  \cite{Gogoladze:2011aa,Ajaib:2013zha}.
{For} this case, $10\%$  or better  $t$-$b$-$\tau$  Yukawa unification consistent with all constraints (including the Higgs boson mass) requires $m_0 \gtrsim 1$ TeV and $m_{1/2} \gtrsim 1$ TeV \cite{Gogoladze:2011aa, Ajaib:2013zha}.
The colored sparticle spectrum  does not change much \cite{Gogoladze:2012ii} if we consider different {mass} relations among {the} gauginos at  $M_{\rm GUT}$, but the sleptons can be light.
Again it leads to heavy  {first and second} generation squarks which are beyond the reach of LHC 14 \cite{cms_lim}.

In both the above mentioned scenarios the sfermions were all assumed to have universal masses at $M_{\rm GUT}$.
The main motivation for this assumption is based on the constraints obtained from flavor-changing neutral current (FCNC) processes \cite{Martin:1997ns}.
It was shown in ref. \cite{Baer:2004xx} that constraints from FCNC processes, for the case when third generation sfermion masses  are split from {masses of the} first and second { {generations}}, are very mild and easily satisfied. {It therefore} allows for significantly {lighter} first two family sfermions, while keeping the third generation sfermions relatively heavy. We adopt
this approach in this paper and we will show that it is possible to have $t$-$b$-$\tau$ YU with LHC accessible {first and second} generation sfermions.

Another motivation for   considering split sfermion {families} is related to  the deviation of the observed muon anomalous magnetic moment
$a_{\mu}=(g-2)_{\mu}/2$ (muon $g-2$) from its SM prediction \cite{Hagiwara:2011af}
\begin{eqnarray}
\label{gg-22}
\Delta a_{\mu}\equiv a_{\mu}({\rm exp})-a_{\mu}({\rm SM})= (28.6 \pm 8.0) \times 10^{-10}.
\end{eqnarray}
If supersymmetry is {to provide a resolution of this} discrepancy, the smuon and gaugino (bino or wino) SSB masses should not be much heavier { {than}} a few {hundred}  GeV. On the other hand, as we mentioned above, $t$-$b$-$\tau$ YU   requires \cite{Baer:2012cp,Gogoladze:2012ii,Gogoladze:2011aa,Ajaib:2013zha} the sleptons to be around a TeV or above, {if} universality among sfermion {masses} is assumed at $M_{\rm GUT}$. Our analysis in the following sections show that the non-universal gaugino case with split family sfermions can {resolve} the $g-2$ discrepancy and also realize $t$-$b$-$\tau$ Yukawa unification, while {staying} consistent with all current experimental data.

We note that recently there {have been} several attempts to {resolve the discrepancy} within the MSSM framework assuming non-universal SSB mass terms at $M_{\rm GUT}$ for gauginos \cite{Akula:2013ioa} or sfermions \cite{Baer:2004xx,Ibe:2013oha}.

The outline for the rest of the paper is as follows. In Section \ref{pheno} we summarize the scanning procedure and the
experimental constraints applied in our analysis. In Sections \ref{so10-nugm} and \ref{so10-ugm} we present the results for supersymmetric SO(10) models {with} non-universal and universal gaugino masses, {respectively}. Tables with benchmark points for both cases are also presented, {and Section \ref{conclusions} contains our conclusions}.

\section{Phenomenological constraints and scanning procedure \label{pheno}}

We employ the ISAJET~7.84 package~\cite{ISAJET}
to perform random scans over the parameter space.
In this package, the weak scale values of gauge and third
generation Yukawa couplings are evolved to
$M_{\rm GUT}$ via the MSSM renormalization group equations (RGEs)
in the $\overline{DR}$ regularization scheme.
We do not strictly enforce the unification condition
$g_3=g_1=g_2$ at $M_{\rm GUT}$, since a few percent deviation
from unification can be assigned to unknown GUT-scale threshold
corrections~\cite{Hisano:1992jj}.
With the boundary conditions given at $M_{\rm GUT}$,
all the SSB parameters, along with the gauge and third family Yukawa couplings,
are evolved back to the weak scale $M_{\rm Z}$.

In evaluating  the Yukawa couplings the SUSY threshold
corrections~\cite{Pierce:1996zz} are taken into account
at a common scale  $M_S= \sqrt{m_{\tilde t_L}m_{\tilde t_R}}$.
The entire parameter set is iteratively run between
$M_{\rm Z}$ and $M_{\rm GUT}$ using the full 2-loop RGEs
until a stable solution is obtained.
To better account for the leading-log corrections, one-loop step-beta
functions are adopted for the gauge and Yukawa couplings, and
the SSB scalar mass parameters $m_i$ are extracted from RGEs at appropriate scales
$m_i=m_i(m_i)$.The RGE-improved 1-loop effective potential is minimized
at an optimized scale  $M_S$, which effectively
accounts for the leading 2-loop corrections. Full 1-loop radiative
corrections are incorporated for all sparticle masses.

We implement the following random scanning procedure: A uniform and logarithmic distribution of random points is first generated in the given parameter space.
The function RNORMX \cite{Leva} is then employed
to generate a Gaussian distribution around each point in the parameter space.  The data points
collected all satisfy
the requirement of radiative electroweak symmetry breaking  (REWSB)   \cite{Ibanez:1982fr},
with the neutralino in each case being the LSP. After collecting the data, we impose
the mass bounds on all the particles \cite{Nakamura:2010zzi} and use the
IsaTools package~\cite{Baer:2002fv}
to implement the various phenomenological constraints. We successively apply the following experimental constraints on the data that
we acquire from  ISAJET~7.84:
\begin{table}[h!]\centering
\begin{tabular}{rlc}
$123~{\rm GeV} \leq  m_h  \leq 127~{\rm GeV}$~~&\cite{:2012gk,:2012gu}&
\\
$ 0.8 \times 10^{-9} \leq BR(B_s \rightarrow \mu^+ \mu^-) $&$ \leq\, 6.2 \times 10^{-9} \;
 (2\sigma)$        &   \cite{:2007kv}      \\
$2.99 \times 10^{-4} \leq BR(b \rightarrow s \gamma) $&$ \leq\, 3.87 \times 10^{-4} \;
 (2\sigma)$ &   \cite{Barberio:2008fa}  \\
$0.15 \leq \frac{BR(B_u\rightarrow
\tau \nu_{\tau})_{\rm MSSM}}{BR(B_u\rightarrow \tau \nu_{\tau})_{\rm SM}}$&$ \leq\, 2.41 \;
(3\sigma)$. &   \cite{Barberio:2008fa}
\end{tabular}\label{table}
\end{table}

\noindent
We also implement the following  mass bounds on the sparticle masses:
\begin{table}[h!]\centering
\begin{tabular}{rlc}
 $m_{\tilde{g}} \gtrsim  1.4~{\rm TeV}~ ({\rm for}~ m_{\tilde{g}}\sim m_{\tilde{q}})$ &~\cite{Aad:2012fqa,Chatrchyan:2012jx}\\
 $m_{\tilde{g}}\gtrsim 1~{\rm TeV}~ ({\rm for}~ m_{\tilde{g}}\ll
m_{\tilde{q}})$ &~\cite{Aad:2012fqa,Chatrchyan:2012jx} \\
$M_A \gtrsim 700~{\rm GeV}$~ $({\rm for}$~ $\tan\beta\simeq 48$). & ~\cite{cms-mA}
\end{tabular}\label{table2}
\end{table}

\noindent
Here $m_{\tilde g}$, $m_{\tilde q}$, $M_A$ {respectively} stand for the gluino, 1st/2nd generation {squarks} and {the} CP odd Higgs boson masses.

\section{SO(10) with non-universal gauginos masses}\label{so10-nugm}

In this section we present the sparticle spectroscopy of {Yukawa unified SO(10) with non-universal gaugino masses } at $M_{\rm GUT}$. {The sfermions of the first and second families are assigned a common } SSB mass term $ m_{16_{1,2}}$, {while} the third generation sfermions have a universal SSB mass term $ m_{16_{3}}$. We also {employ} universal SSB mass term for the MSSM Higgs bosons, $m^2_{H_{u}}= m^2_{H_{d}} \equiv m^2_{10}$.  As mentioned earlier, gauge coupling unification is one of the nice features of low scale supersymmetry and  indicates that
the SM gauge symmetry is {embedded within a} a simple gauge group with rank $\geq$ 4. In this case the MSSM gauginos are contained within a single vector  multiplet. {To retain gauge coupling unification  and at the same time have non-universal gaugino masses at $M_{\rm GUT}$, one way is to employ~\cite{Martin:2009ad} non-singlet $F$-terms, compatible with the underlying GUT.} Non-universal gauginos can also be generated  from the $F$-term with a linear combination of two distinct fields of different dimensions \cite{Martin:2013aha}. We can also {consider} two {distinct} sources for supersymmetry breaking \cite{Anandakrishnan:2013cwa}. Since there are many possibilities {for realizing} non-universal
gaugino {masses with either} fixed or arbitrary mass ratios, we {employ} independent masses for {the three} MSSM  gauginos in our study. In this case our analysis will cover {a variety of} scenarios with {non-universal gaugino masses and} split sfermion {families} in the {presence} of $t$-$b$-$\tau$ Yukawa unification. 

{We} have performed random scans in the fundamental parameter space of the model with ranges of the parameters given as follows:
\begin{align}
0 \leq  m_{16_{1,2}}  \leq 1\, \rm{TeV} \nonumber  \\
0 \leq  m_{16_{3}}  \leq 5\, \rm{TeV} \nonumber  \\
-1 \leq  M_{1}  \leq 0\, \rm{TeV} \nonumber  \\
-1 \leq  M_{2}  \leq 0\, \rm{TeV} \nonumber  \\
0 \leq  M_{3}  \leq 5\, \rm{TeV} \nonumber  \\
-3 \leq A_{0}/m_{3}  \leq 3 \nonumber  \\
35 \leq  \tan\beta  \leq 55 \nonumber \\
0 \leq  m_{10}  \leq 5\, \rm{TeV} \nonumber \\
\mu < 0
\label{parameterRange}
\end{align}

\begin{figure}[t!]
\subfigure{\includegraphics[scale=1]{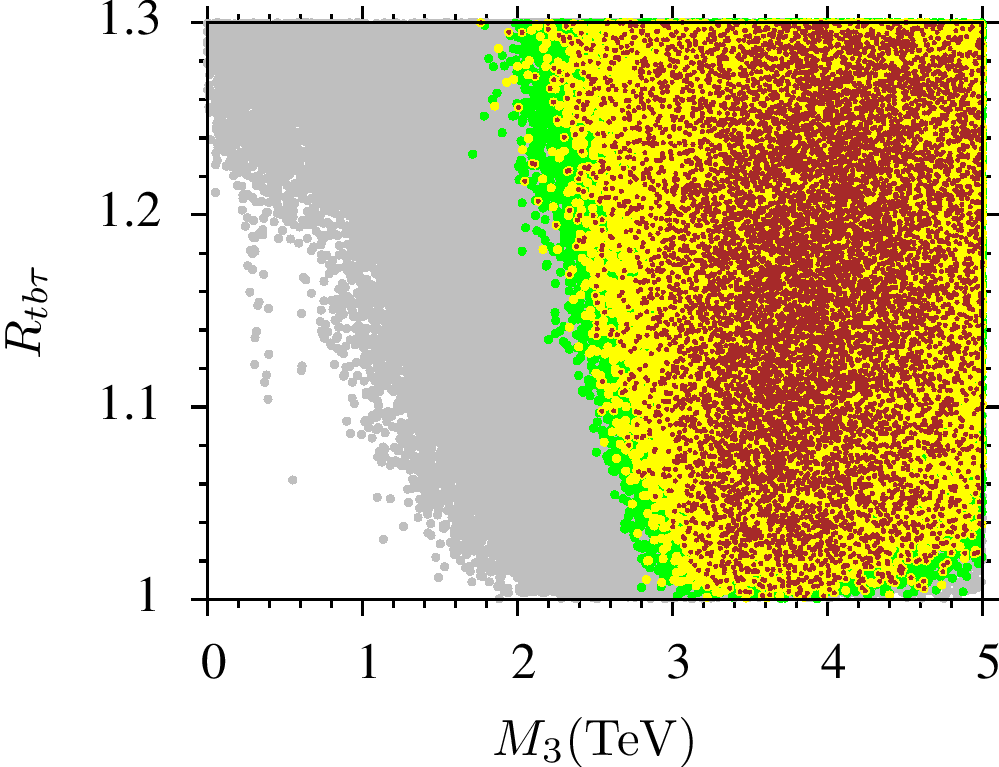}}%
\subfigure{\includegraphics[scale=1]{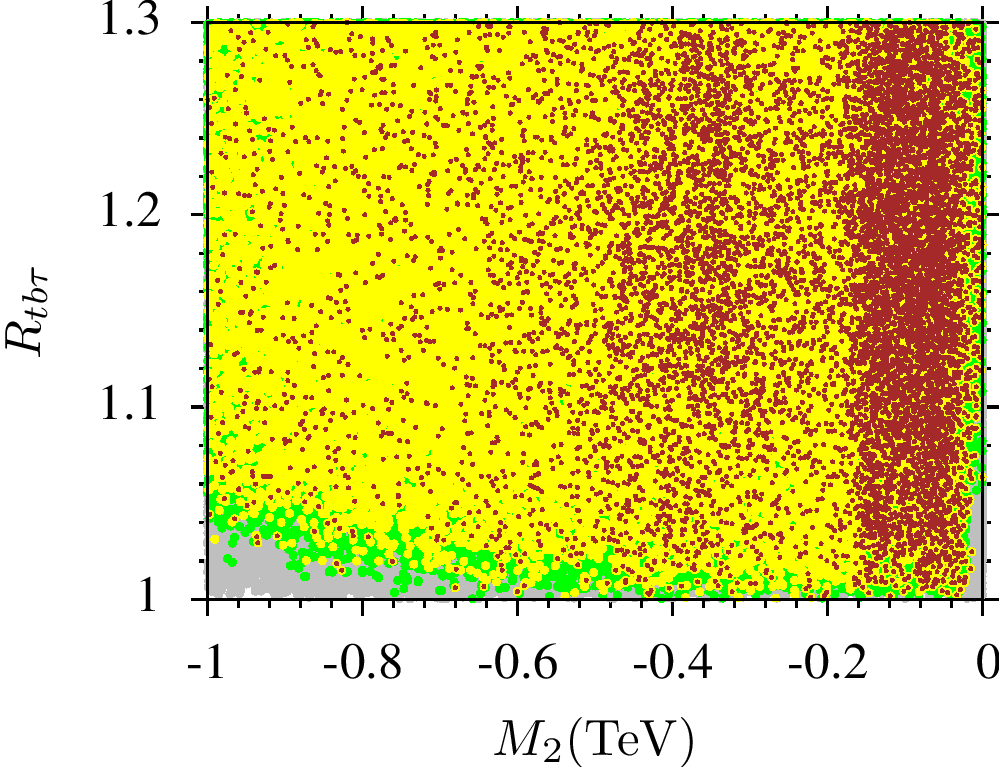}}
\subfigure{\includegraphics[scale=1]{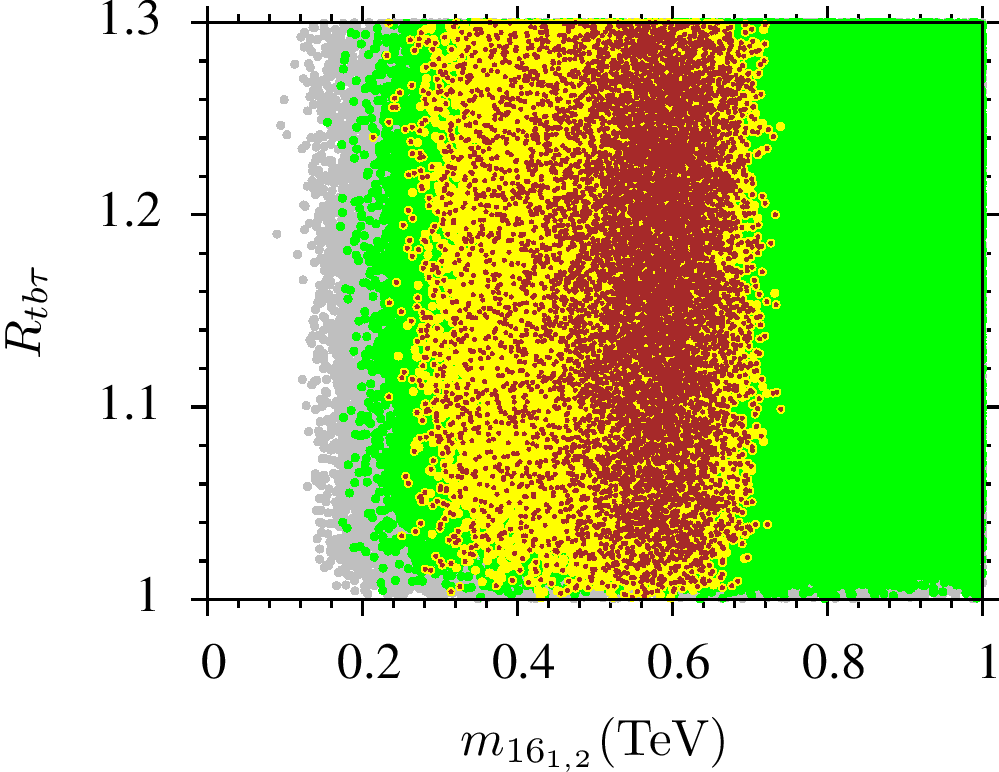}}
\subfigure{\hspace{0.3cm}\includegraphics[width=6.5cm,height=5.3cm]{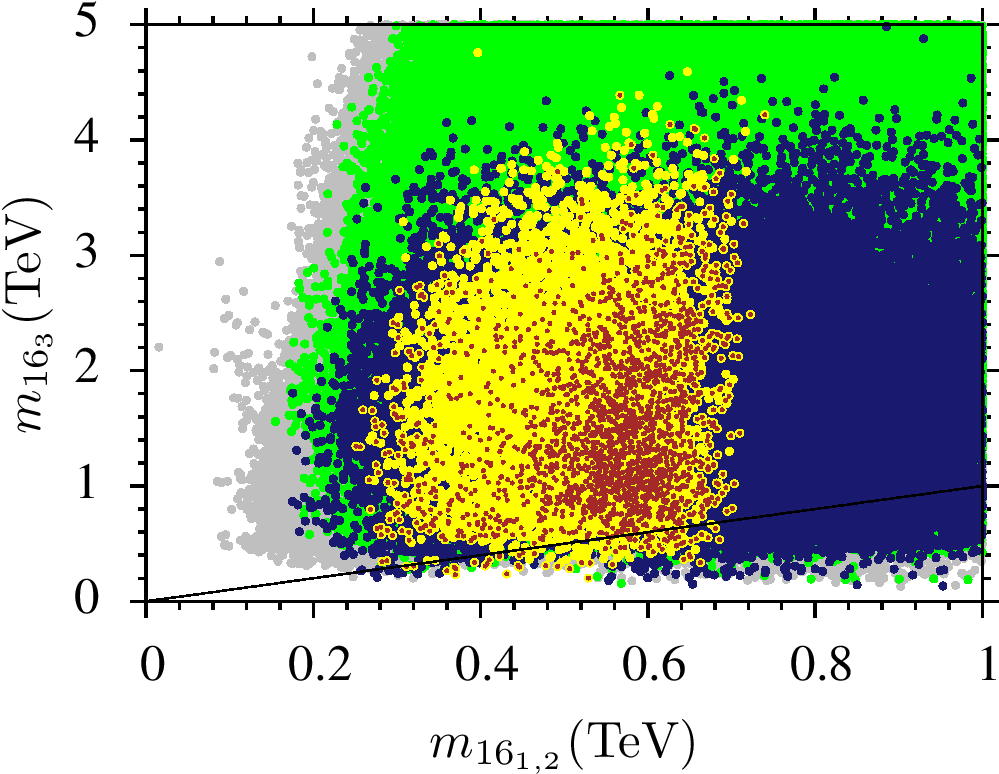}}
\caption{Plots in the $R_{tb\tau} - M_{3}$, $R_{tb\tau} - M_{3}$, $R_{tb\tau} - m_{16_{1,2}}$ and
$m_{16_{3}} - m_{16_{1,2}}$  planes.
{\it Gray} points are consistent with REWSB  and neutralino LSP.  {\it Green} points form a subset of the {\it gray} {points}
and satisfy the sparticle and Higgs mass bounds, {as well as} all other constraints described in Section \ref{pheno}.
{\it Yellow} points are {a subset} of the {\it green} points {and} satisfy the $\Delta a_{\mu}$ constraint in
 Eq. (\ref{gg-22}). {\it Brown} points belong to a subset of {\it yellow} points and satisfy bound on the LSP neutralino relic abundance,  $0.001 \leq \Omega h^2 \leq 1$. In the $m_{16_{3}} - m_{16_{1,2}}$  panel, in  addition, {\it blue} points are a subset of the {\it green} {ones}
 and satisfy $R_{tb\tau}< 1.1$ and {\it yellow}.
The  {\it yellow} region
  is a subset of the blue region, while {\it brown} is subset of {\it yellow} with the definition of the colors mentioned above.}
\label{fig-1}
\end{figure}

\begin{figure}[]
\centering
{\label{fig:2}{\includegraphics[scale=1]{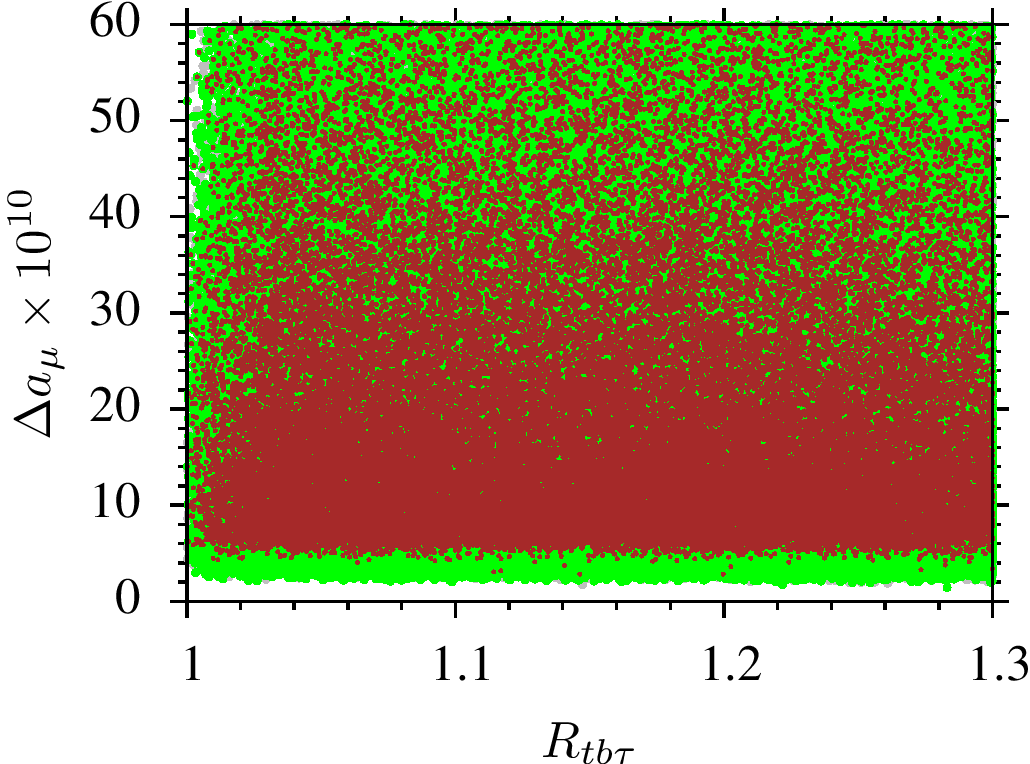}}}
{\label{fig:2}{\includegraphics[scale=1]{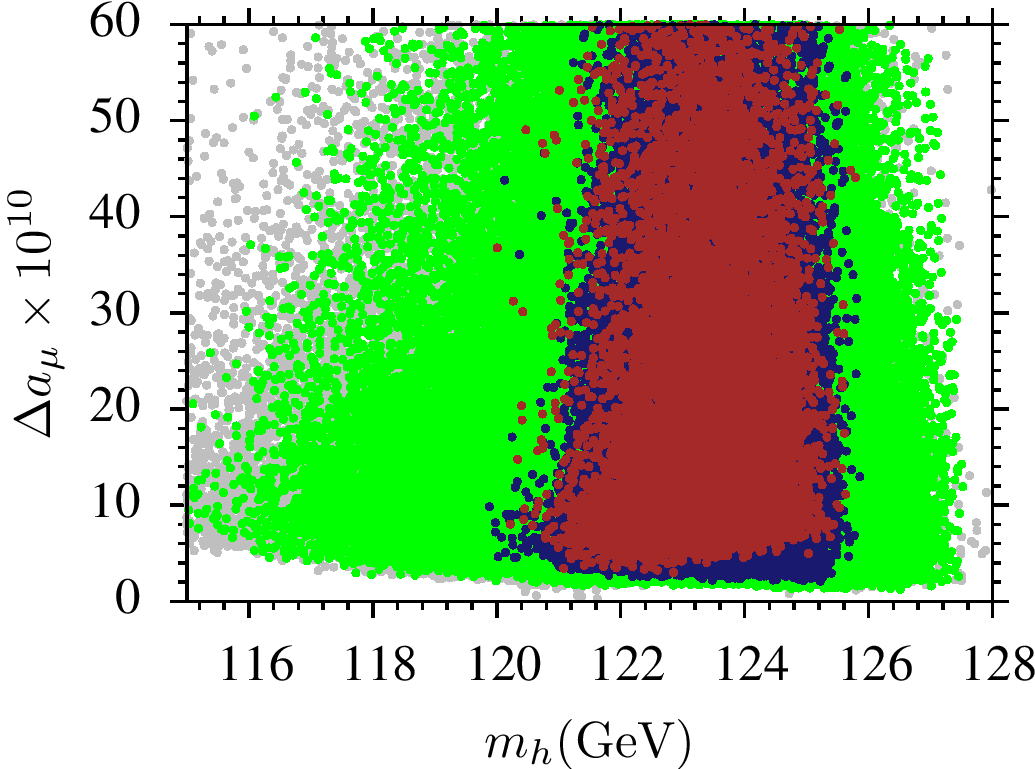}}}
\caption{Plots in the $\Delta  a_{\mu} - R_{tb\tau}$ and $\Delta  a_{\mu} - m_h$ planes.
{\it Gray} points are consistent with REWSB  and neutralino LSP.  {\it Green} points form a subset of the {\it gray} {points}
and satisfy the sparticle and Higgs mass bounds, {as well as} all other constraints described in Section \ref{pheno}.
{\it Brown} points belong to a subset of {\it green} points and satisfy the bound for LSP relic abundance,  $0.001 \leq \Omega h^2 \leq 1$. In the $\Delta  a_{\mu} - m_h$ panel  {\it blue} points are subset of the {\it green} {ones} and satisfy $R_{tb\tau}< 1.1$.
{\it Brown} points are subset of {\it blue} points and have the same definition as above.}
\label{fig-2}
\end{figure}


\begin{figure}[]
\centering
{\label{fig:3}{\includegraphics[scale=1]{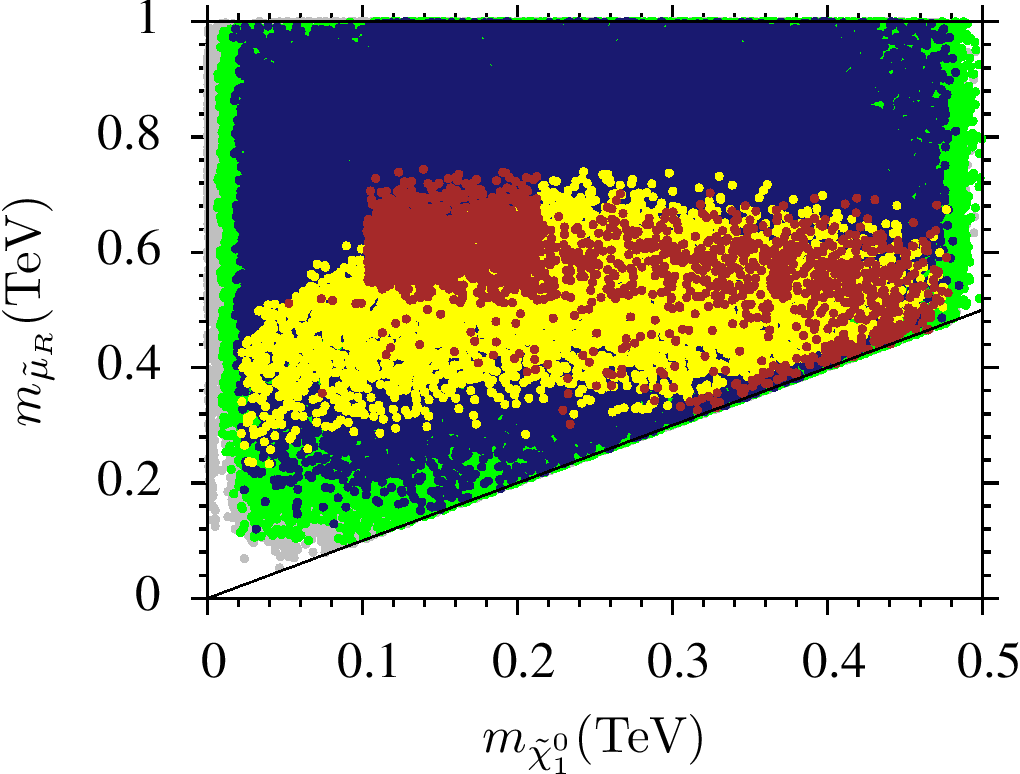}}}
{\label{fig:3}{\includegraphics[scale=1]{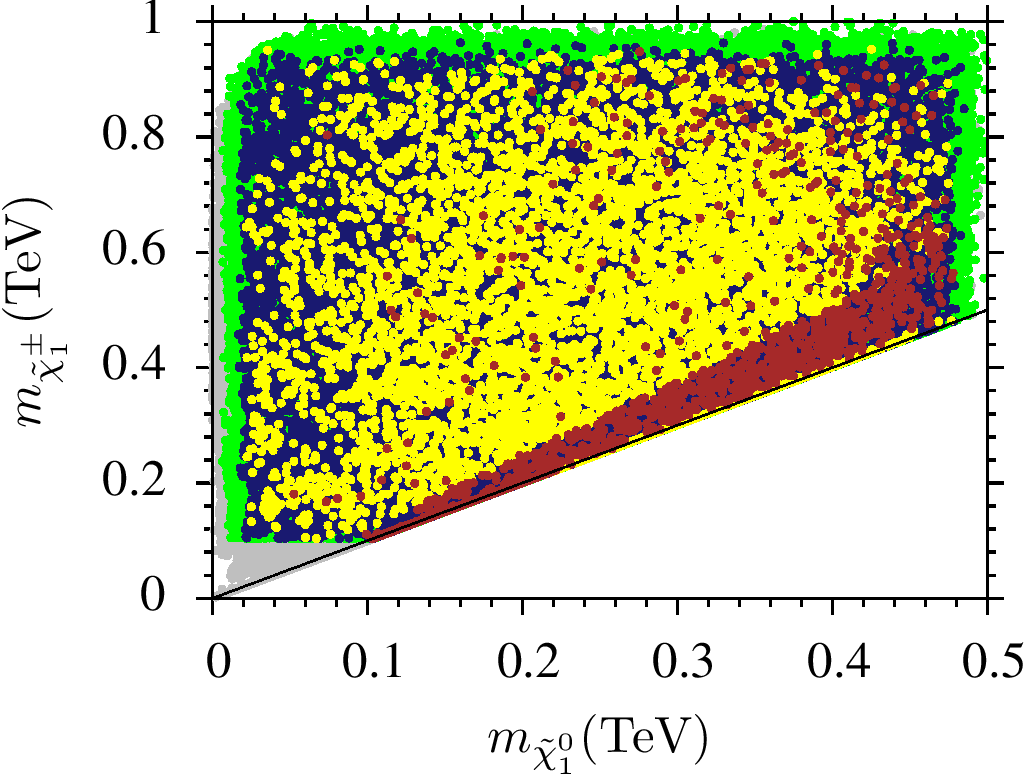}}}
{\label{fig:3}{\includegraphics[scale=1]{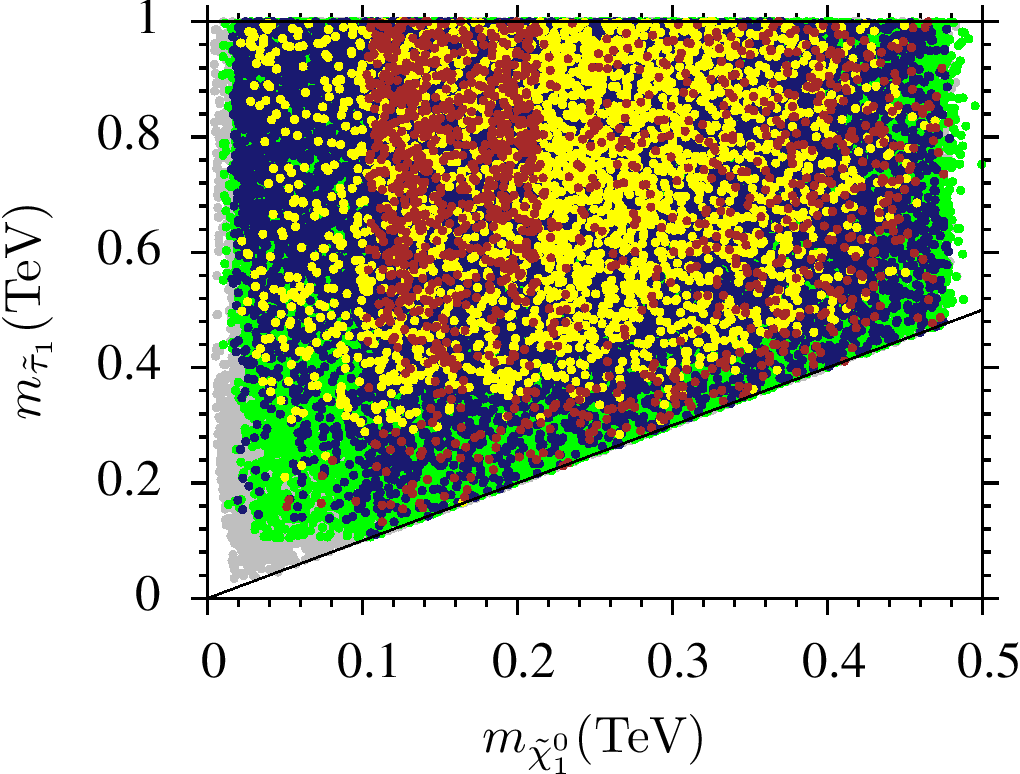}}}
{\label{fig:3}{\includegraphics[scale=1]{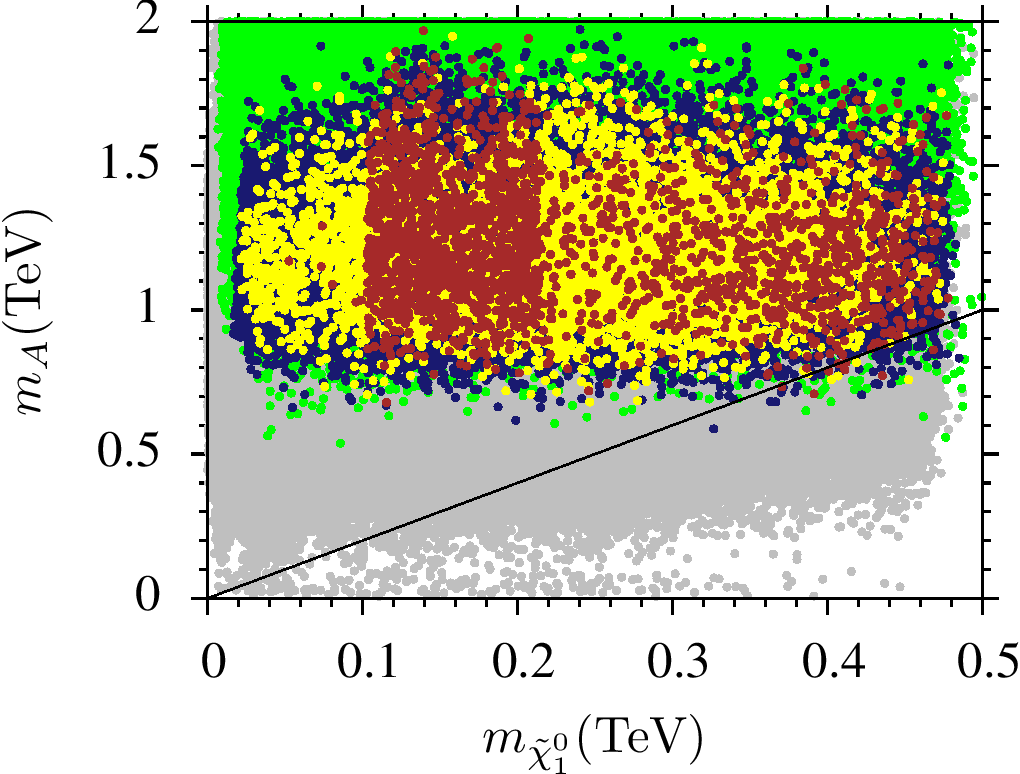}}}
{\label{fig:3}{\includegraphics[scale=1]{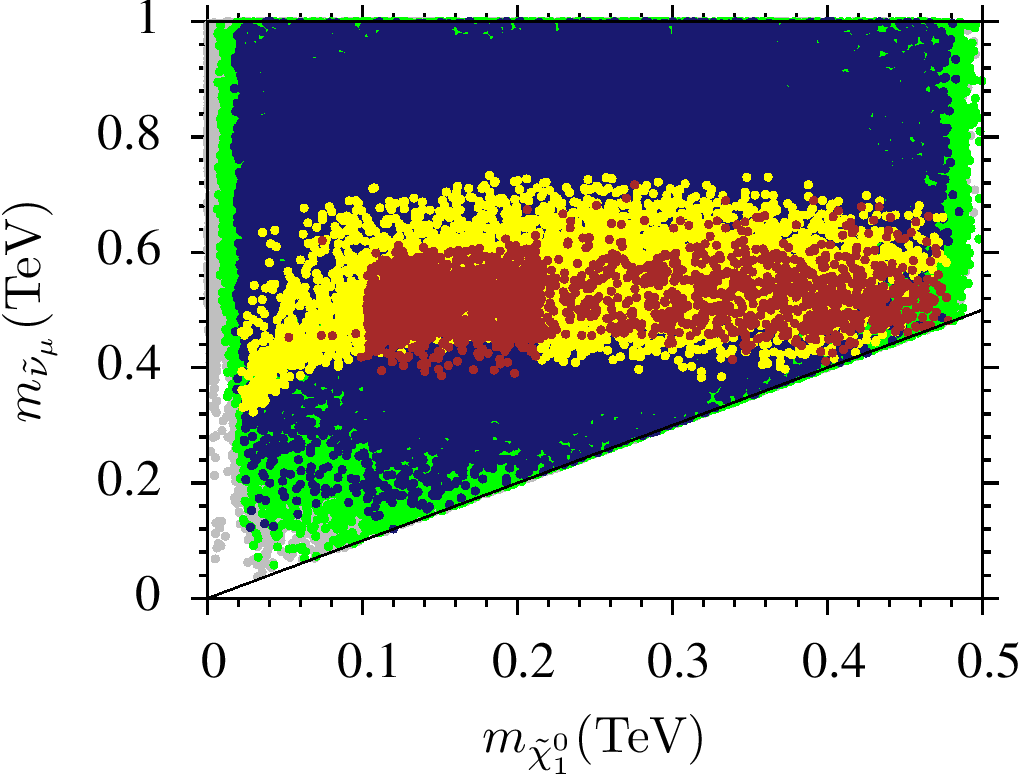}}}
{\label{fig:3}{\includegraphics[scale=1]{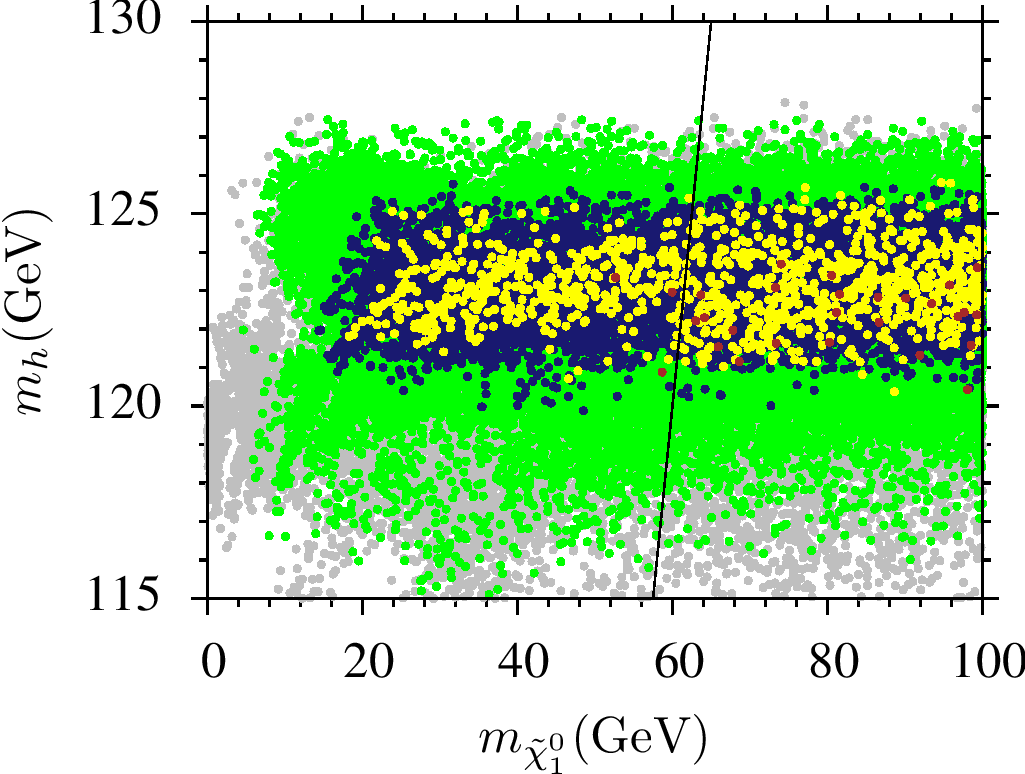}}}
\caption{Color coding  same as in the $m_{16_{3}} - m_{16_{1,2}}$  panel in Figure 1.}
\label{fig-3}
\end{figure}


\begin{figure}[]
\centering
{\label{fig:4}{\includegraphics[scale=1]{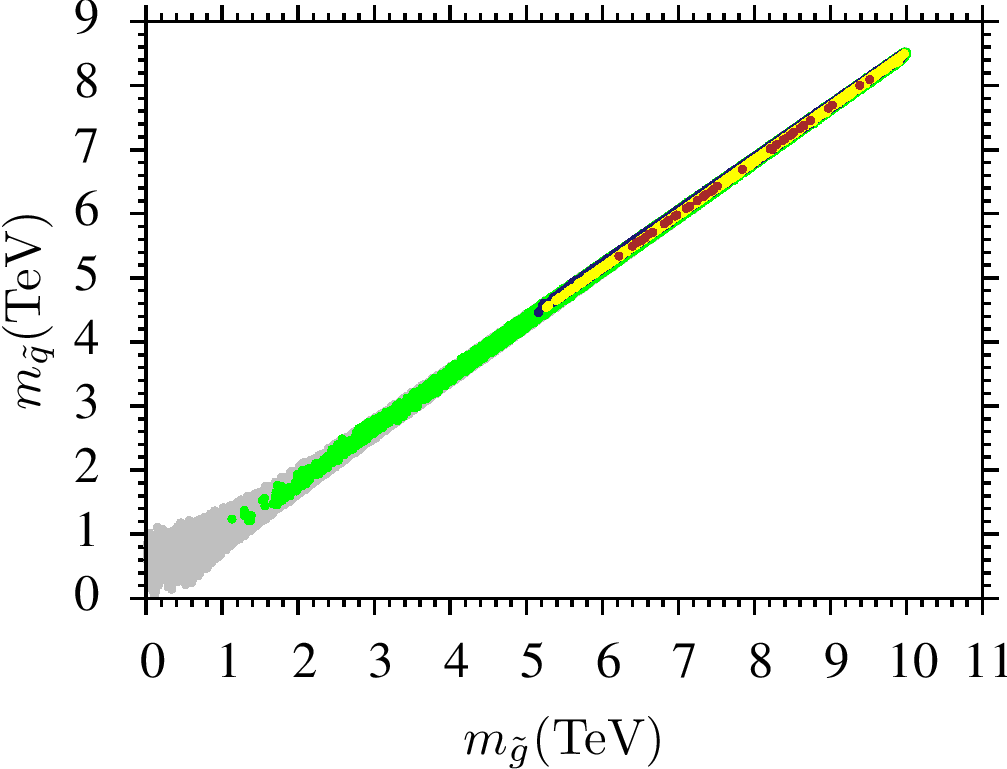}}}
\caption{Color coding  same as in the $m_{16_{3}} - m_{16_{1,2}}$  panel in Figure 1.}
\label{fig-4}
\end{figure}


 In Figure 1 we show the results in the
$R_{tb\tau} - M_{3}$, $R_{tb\tau} - M_{2}$, $R_{tb\tau} - m_{16_{1,2}}$ and
$m_{16_{3}} - m_{16_{1,2}}$  planes.
{\it Gray} points are consistent with REWSB  and neutralino LSP.  {\it Green} points form a subset of the {\it gray} {ones}
and satisfy sparticles and Higgs mass bounds and all other constraints described in Section \ref{pheno}.
{\it Yellow} points are { {a}} subset of the {\it green} points satisfy {the muon $g-2$ constraint   given in
Eq. (\ref{gg-22})}. {\it Brown} points {are} a subset of {\it yellow} points and satisfy the following neutralino relic abundance {constraint},  $0.001 \leq \Omega h^2 \leq 1$. We have chosen to display our  results for a wider range of $\Omega h^2$ keeping in mind that one can always find points
which are compatible with the current WMAP range for relic abundance  with dedicated scans within the {\it brown} regions. In the $m_{16_{3}} - m_{16_{1,2}}$  plane, in  addition, {the} {\it blue} points are a subset of the {\it green} {ones} and satisfy $R_{tb\tau}< 1.1$. In this panel {the} {\it yellow} region  is a subset of the blue, and {\it brown} is a subset of the {\it yellow} region with the {color} definitions {the} same as mentioned above.

From the $R_{tb\tau} - M_{3}$ plane we see that just from the REWSB condition ({\it gray} points), we cannot have  $M_3 \lesssim 1$ TeV (or equivalently $m_{\tilde g} \lesssim 3$ TeV) {if} we  demand $t$-$b$-$\tau$ Yukawa unification better than $10\%$.  The reasons for such a heavy gluino mass are the {combined} effects from REWSB and the necessity for {appropriate} threshold corrections for  $t$-$b$-$\tau$ Yukawa unification \cite{Ajaib:2013zha}.  {If} we apply the current experimental { {constraint}}, the lower mass bound on the gluino   changes drastically ({\it green} points). {In particular}, $t$-$b$-$\tau$ Yukawa unification better than $10\%$ {requires that} ${M_3} \gtrsim 2.5$ TeV.
This bound is mostly dictated from the Higgs mass constraint ($123~{\rm GeV} \leq m_h \leq 127~{\rm GeV}$), the reason {being} that
 {for}  $t$-$b$-$\tau$ Yukawa unification {with} non-universal and opposite sign  gaugino {masses}, the following condition is usually satisfied: $A_t/ M_{S} <  1$   \cite{Ajaib:2013zha}. On the other hand, it is known \cite{Carena:2002es} that the light CP even Higgs boson mass
receives significant contribution from the $A_t$ term if $A_t/M_S \gtrsim 1$. We can  therefore conclude that there is no significant contribution from finite corrections to the CP even Higgs boson mass if we require almost perfect Yukawa unification and the Higgs mass is mostly generated from logarithmic corrections involving the stop quark. It was also shown in \cite{Ajaib:2013zha} that the stop quark { {mass}} {in this case} has to be $\gtrsim 5$ TeV in order to satisfy the Higgs mass bound.  Another constraint from Yukawa unification, namely, $M_{3}> m_{16_{3}}$, implies that the stop quark mass is mostly determined from radiative corrections from the gluino.  This, therefore, is the reason why the gluino mass affects the Higgs mass bound so strongly for $t$-$b$-$\tau$ Yukawa unification {better} than 10\%. {The} {\it yellow} points show that in this scenario supersymmetry can easily provide the {desired} contribution to the muon $g-2$ anomaly.  We will show later that there are several channels that can generate the  correct relic abundance for neutralino dark matter, {displayed by the }{\it brown} points.

In the $R_{tb\tau} - M_{2}$ plane we observe a very mild constraint on {the parameter} $M_2$ from  $t$-$b$-$\tau$ Yukawa unification and all current experimental data including muon $g-2$ anomaly and dark matter relic abundance.  The same conclusion applies to $M_1$  which is the reason we do not display results in terms of $M_1$ here.

Since the Higgs mass bound and  $t$-$b$-$\tau$ Yukawa unification condition only affects the third generations squarks, {the first and second} generation sfermions can be as light as 100 GeV, as seen from the $R_{tb\tau} - m_{16_{1,2}}$ plane ({\it gray} points). The current experimental data (including the limit $m_{\tilde q} \gtrsim 1.5$  TeV) does not {significantly} change the lower bound on $ m_{16_{1,2}}$. {The reason is that here the gluino and wino masses are independent of each other and a large gluino mass ($m_{\tilde g} > 5$ TeV) automatically pushes the squark masses to a few TeV despite the low initial value $m_{16_{1,2}}$ at  $M_{\rm GUT}$.} This allows for low values of $ m_{16_{1,2}}$ while still being consistent with all experimental results.
After applying the muon $g-2$ constraint we obtain $0.3~{\rm TeV} \lesssim m_{16_{1,2}} \lesssim 0.7~{\rm TeV}$. Again, {\it brown} points show that
the correct relic abundance {is} easily achieved once the muon $g-2$ constraint is applied.

From the $m_{16_{3}} - m_{16_{1,2}}$  plane we learn that it is possible
to have solutions consistent with muon $g-2$ anomaly when $m_{16_{3}} = m_{16_{1,2}}$ with arbitrary and opposite sign gaugino mass ratios at $M_{\rm GUT}$ (see {\it yellow} points which are subset  of {\it blue} points showing $10\%$ or better $t$-$b$-$\tau$ Yukawa unification).  We find that in this case $M_{3}> 3 \cdot M_{2}$ needs to be satisfied at $M_{\rm GUT}$. We also see that the solution consistent with muon $g-2$ anomaly mostly occurs {for} $m_{16_{3}} > m_{16_{1,2}}$.

\begin{table}[]\hspace{-1.0cm}
\centering
\begin{tabular}{|c|ccccc|}
\hline
\hline
                 & Point 1 & Point 2 & Point 3 & Point 4 & Point 5\\

\hline
$m_{16_{1,2}}$        & 375.9 & 353.2 & 639 & 450.7 & 620.6 \\
$m_{16_{3}}$          & 2257 & 562.9 & 3113  & 1634 & 3580 \\
$M_{1} $       & -981.9 & -821.8 & -526.8 & -739.2 & -699.1 \\
$M_{2} $       & -701.8 & -640.6 & -266.5 & -389.9 & -372.4 \\
$M_{3} $       & 4299 & 3589 & 4305  & 3685 & 4771 \\
$\tan\beta$      & 51.5 & 51.1 & 50.7 & 51.4 & 50.7 \\
$A_0/m_{16_{3}}$      & -2.06 & -0.73 & -1.62 & 1.87 & 1.76 \\
$m_{10}$          & 2512 & 988.7 & 370.5 & 189.4 & 1315 \\
$m_t$            & 173.3  & 173.3 & 173.3 & 173.3 & 173.3 \\
\hline
& & & & & \\
$\mu$ & -4845 & -3707 & -5648 & -3926 & -6276\\
$\Delta a_{\mu}$  & $ \mathbf{ 31.5\times 10^{-10} } $ & $ \mathbf{31.9\times 10^{-10} } $ & $ \mathbf{ 25.7\times 10^{-10} } $ & $ \mathbf{34.8 \times 10^{-10} } $ & $ \mathbf{28.6 \times 10^{-10} } $ \\ & & & & & \\

\hline
$m_h$            & {124.6} & {123.3} & {124.7} & {123.1} & {125} \\
$m_H$            & 1280 & 1244 & 1293  & 778.5 & 1651\\
$m_A$            & 1272 & 1236 & 1285 & {\bf 773.5} & 1641 \\
$m_{H^{\pm}}$    & 1284 & 1248 & 1297 & 784.7 & 1654 \\

\hline
$m_{\tilde{\chi}^0_{1,2}}$
                 & {\bf 466.3}, 680.9 & {\bf 392.9}, 625.9 & {\bf 259.9}, {\bf  304.8} & {\bf 361.4}, 426.6 & {\bf 340.1}, {\bf 402.7} \\

$m_{\tilde{\chi}^0_{3,4}}$
                 & 4843, 4843 & 3709, 3710 & 5635, 5635 & 3930, 3930 & 6262, 6262\\

$m_{\tilde{\chi}^{\pm}_{1,2}}$
                 & 683.5, 4841 & 627.8, 3710 & {\bf 306.1}, 5633 & 428.3, 3929 & {\bf 404.4}, 6259 \\

$m_{\tilde{g}}$  & 8599 & 7247 & 8650 & 7436 & 9517\\
\hline $m_{ \tilde{u}_{L,R}}$
                 & 7332, 7346 & 6204, 6214  & 7373, 7401 & 6351, 6372 & 8094, 8120\\
$m_{\tilde{t}_{1,2}}$
                 & 6232, 6419 & 5363, 5496 & 6545, 6730 & 5734, 5830 & 7157, 7375\\
\hline $m_{ \tilde{d}_{L,R}}$
                 & 7332, 7348 & 6204, 6216 & 7374, 7406 & 6351, 6374 & 8094, 8125\\
$m_{\tilde{b}_{1,2}}$
                 & 6352, 6425 & 5427, 5489 & 6659, 6743 & 5775, 5837 & 7316, 7412\\
\hline
$m_{\tilde{\nu}_{1,2}}$
                 & 521.2 & 503.4 & 568 & 456.2 & 555.1\\
$m_{\tilde{\nu}_{3}}$
                 & 1991 & 778.7 & 2846 & 1437 & 321.6\\
\hline
$m_{ \tilde{e}_{L,R}}$
                & 546.5, {493.7} & 519.1, 448.8 & 592.3, 642.6 & 470.9, 510.2 & 588.4, 640.2 \\
$m_{\tilde{\tau}_{1,2}}$
                & 1469, 1994  & {437.1}, 909.9 & 2454, 2846  & 1012, 1446 & 2695, 3209\\
\hline

$\sigma_{SI}({\rm pb})$
                & $0.29\times 10^{-11}$ & $ 0.49\times 10^{-11} $ & $ 0.22\times 10^{-11} $ & $0.16\times 10^{-10} $
& $0.13\times 10^{-11} $\\

$\sigma_{SD}({\rm pb})$
                & $0.10\times 10^{-9}$ &$ 0.33\times 10^{-9} $ & $ 0.46\times 10^{-10} $ & $0.28\times 10^{-9} $
& $0.29\times 10^{-10} $\\

$\Omega_{CDM}h^{2}$
                &  0.12 & 0.12 & 0.11 & 0.11 & 0.13 \\
\hline

$R$             & 1.06 &  1.05 & 1.05 & 1.01 &  1.09  \\

\hline
\hline
\end{tabular}
\caption{Benchmark points with $\Delta a_{\mu} $ within $1\sigma$ deviation from its theoretical value. All the masses are in units of GeV. Points are chosen to be consistent with all the constraints described on Section \ref{pheno}. Point 1 depicts a solution for smuon (selectron) coannihilation, while point 2 represents stau-coannihilation. Points 3 and 4 display chargino-neutralino coannihilation and A-resonance solutions, respectively. Point 5 shows a solution with a 125 GeV Higgs boson and the central value of muon $g-2$.}
\label{table:1}
\end{table}

In Figure 2 we show the results  in the $\Delta  a_{\mu} - R_{tb\tau}$ and $\Delta  a_{\mu} - m_h$ planes.
{\it Gray} points are consistent with REWSB  and neutralino LSP.  {\it Green} points form a subset of the {\it gray} {points}
and satisfy the sparticles, Higgs mass bound and all other constraints described in Section \ref{pheno}.
{\it Brown} points {are a} subset of the {\it green} points and satisfy the following bound on the neutralino relic abundance:  $0.001 \leq \Omega h^2 \leq 1$. In the $\Delta  a_{\mu} - m_h$ panel, the {\it blue} points are a subset of the {\it green} {ones}  and satisfy $R_{tb\tau}< 1.1$.
In this plane the {\it brown} points are a subset of the {\it blue} {ones} {with the} same definition mentioned above. We can see from the $\Delta  a_{\mu} - R_{tb\tau}$ plane that a notable region of the parameter space simultaneously yields perfect $t$-$b$-$\tau$ Yukawa unification along with the {desired} contribution to the muon $g-2$ anomaly, while satisfying all experimental constraints described in in Section \ref{pheno}.

The $\Delta  a_{\mu} - m_h$ panel  shows that it is possible to have a 125 GeV light CP-even Higgs boson consistent
with the {desired} contribution to the muon $g-2$ anomaly. {The desired contribution to the muon $g-2$ anomaly and a 125 GeV Higgs cannot be easily attained for a broad class of low scale supersymmetric model}. For instance, it was shown in \cite{Okada:2013ija} that with universal SSB gaugino and sfermion masses at $M_{\rm GUT}$, it is very hard to simultaneously have a 125 GeV Higgs boson mass and the {desired} $\Delta a_{\mu}$ { {within}} 1$\sigma$ deviation from its theoretical value. In our case this is easily achieved {and is { {also}} compatible with} good $t$-$b$-$\tau$ Yukawa unification ({\it blue} points).

In Figure 3 we present our results in $m_{\tilde \mu_R} - m_{\tilde \chi_{1}^{0}}$, $m_{\tilde \chi_{1}^{\pm}} - m_{\tilde \chi_{1}^{0}}$,
$m_{\tilde \tau_1} - m_{\tilde \chi_{1}^{0}}$, $m_{A} - m_{\tilde \chi_{1}^{0}}$, $m_{\tilde \nu_{\mu}} - m_{\tilde \chi_{1}^{0}}$ and
$m_{h} - m_{\tilde \chi_{1}^{0}}$ planes in order to show the different channels contributing to yield the correct neutralino dark matter relic abundance.  We see that all the channels are consistent  with the {desired} contribution to  muon $g-2$ anomaly.  We also observe that the slepton mass in this scenario can be around 200 GeV, {and so} there is hope that it can be tested at the LHC. The results in the $m_{\tilde \chi_{1}^{\pm}} - m_{\tilde \chi_{1}^{0}}$ plane {exhibit} bino-wino and bino-higgsino mixed dark matter scenarios. The
$m_{h} - m_{\tilde \chi_{1}^{0}}$ {panel} shows the presence of light Higgs and $Z$-resonance neutralino dark matter solutions, consistent with Yukawa unification. The solid  line in this plane stands for the relation $m_{h}= 2\cdot  m_{\tilde \chi_{1}^{0}}$.
Finally, the $m_{\tilde q} - m_{\tilde g}$ panel in Figure 4 shows that $t$-$b$-$\tau$ Yukawa unification predicts $m_{\tilde q}\gtrsim 4$ TeV and
 $m_{\tilde g}\gtrsim 5$ TeV ({\it blue} points), which may be difficult to observe at LHC 14.

 Table \ref{table:1} lists four benchmark points for this scenario that have good Yukawa unification, satisfy the Higgs mass bound, yield the desired $\Delta a_{\mu}$, and satisfy all other constraints described in section \ref{pheno}. In addition, the relic density is within the WMAP limit on the dark matter abundance. Point 1 depicts a solution for smuon/selectron-coannihilation while point 2 represents stau-coannihilation. Points 3 and 4 display chargino-neutralino coannihilation and A-resonance ($m_A \sim 2 m_{\chi^0_1}$) solutions, respectively.

\section{SO(10) with  universal gauginos masses}\label{so10-ugm}

In this section we present the SO(10) sparticle spectroscopy {corresponding to} $t$-$b$-$\tau$ Yukawa unification {and universal} gaugino mass terms at $M_{\rm GUT}$.  As shown in ref. \cite{Blazek:2001sb}, in this case we must have non-universal SSB mass$^2$ terms for {the} MSSM Higgs bosons, {namely} $m^2_{H_{u}} \neq m^2_{H_{d}}$ at $M_{\rm GUT}$. Otherwise, it is very difficult to {simultaneously implement} REWSB and $t$-$b$-$\tau$ Yukawa unification  (for discussion see ref. \cite{Ajaib:2013zha}).
{As in the previous case}, the sfermions from {the first and second families} have common universal SSB mass terms $m_{16_{1,2}}$, and the third generation sfermions have {the} universal SSB mass term $m_{16_{3}}$.

{The} random scans {are performed for} the following range of {parameters}:
\begin{align}
0 \leq  m_{16_{1,2}}  \leq 1\, \rm{TeV} \nonumber  \\
0 \leq  m_{16_{3}}  \leq 5\, \rm{TeV} \nonumber  \\
0 \leq  M_{1/2}  \leq 2\, \rm{TeV} \nonumber  \\
-3 \leq A_{0}/m_{3}  \leq 3 \nonumber  \\
35 \leq  \tan\beta  \leq 55 \nonumber \\
0 \leq  m_{H_{u}}  \leq 30\, \rm{TeV} \nonumber \\
0 \leq  m_{H_{d}}  \leq 30\, \rm{TeV} \nonumber \\
\mu > 0.
\label{para-2}
\end{align}

\begin{figure}[]
\centering
{\label{fig:5}{\includegraphics[scale=0.4]{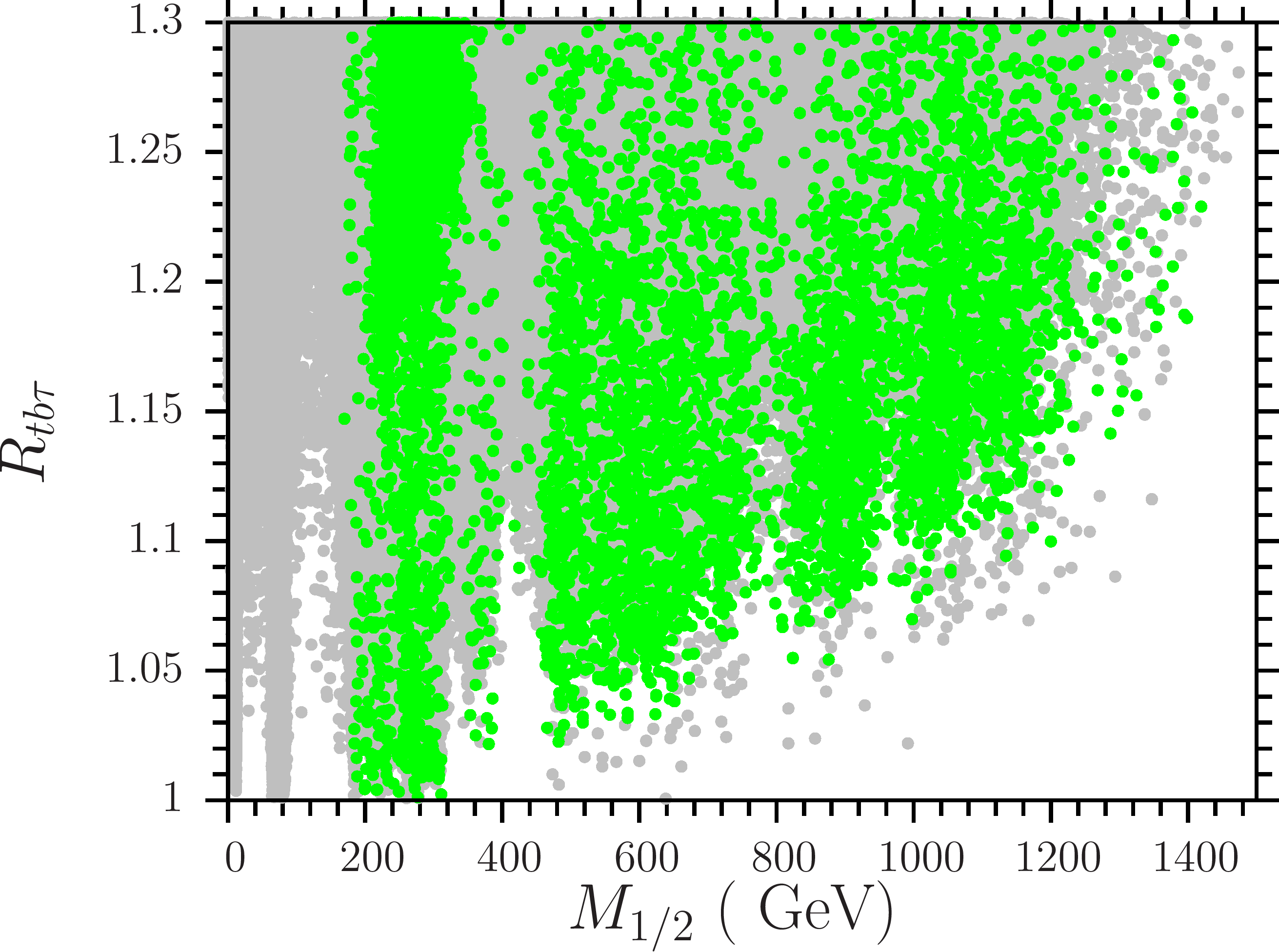}}}
{\label{fig:5}{\includegraphics[scale=0.4]{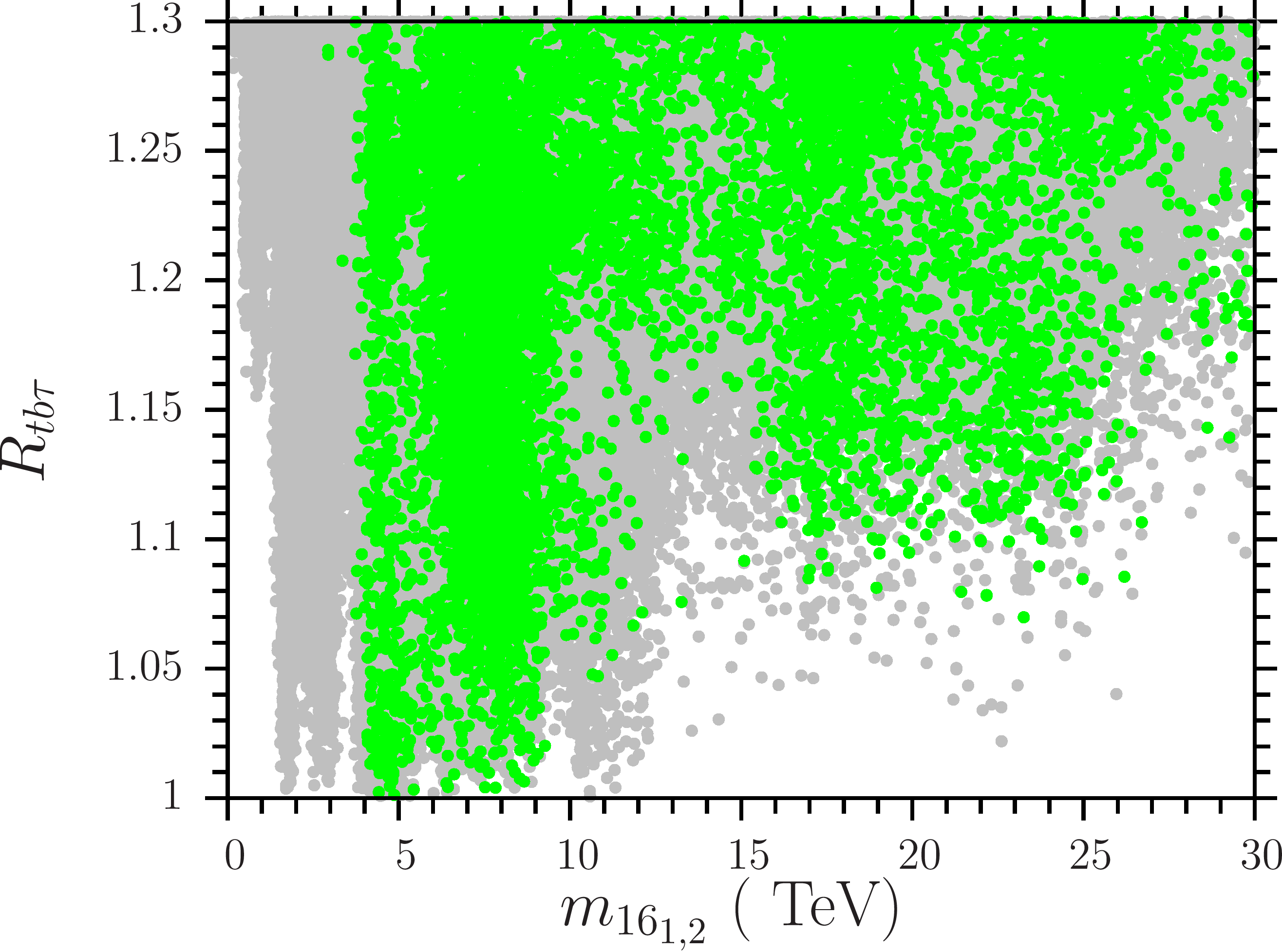}}}
{\label{fig:5}{\includegraphics[scale=0.4]{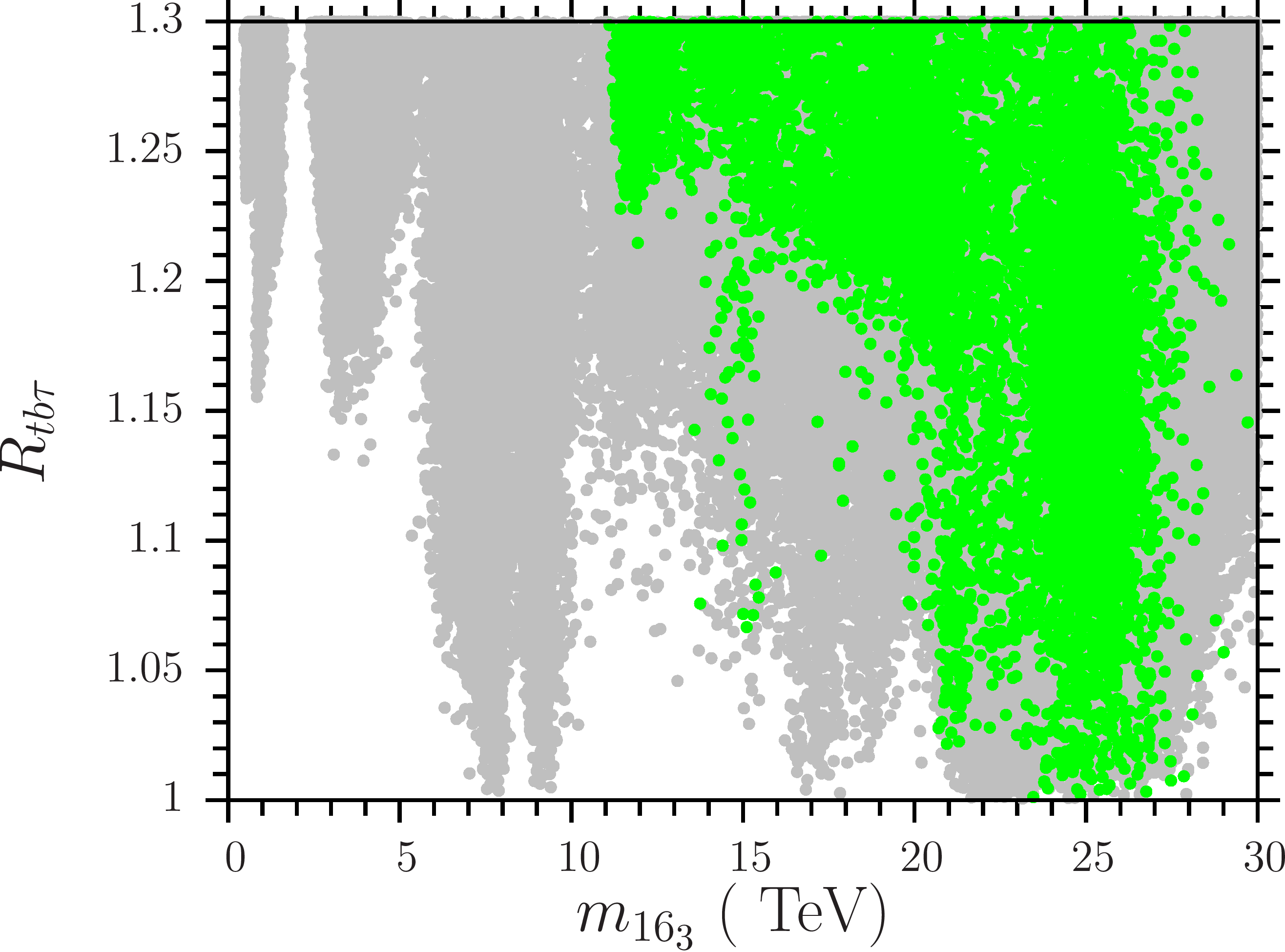}}}
{\label{fig:5}{\includegraphics[scale=0.4]{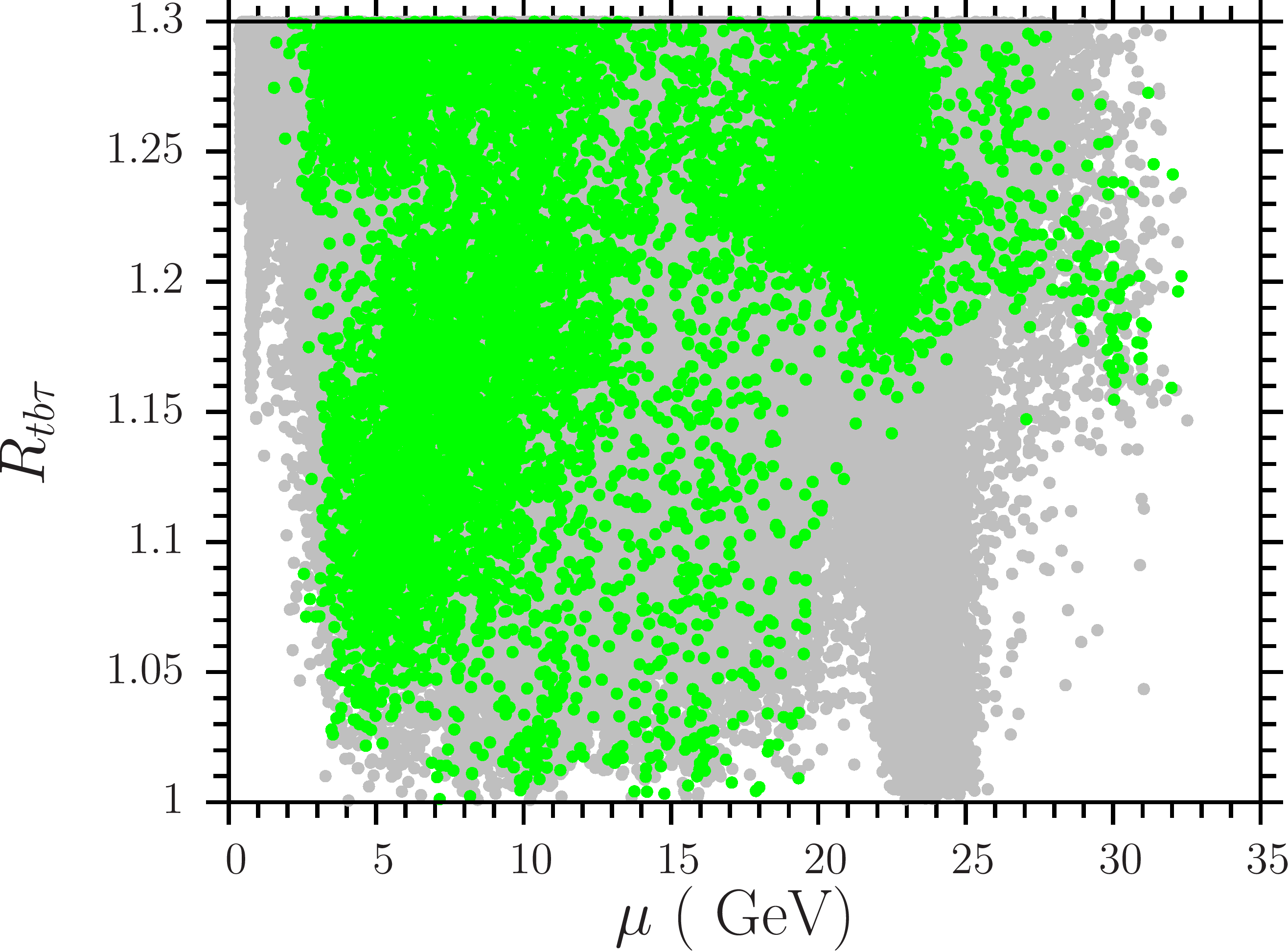}}}
\caption{Plots in the $R_{tb\tau} - M_{1/2}$,  $R_{tb\tau} - m_{16_{1,2}}$, $R_{tb\tau} - m_{16_{3}}$ and
$m_{16_{3}} - \mu$  planes.
{\it Gray} points are consistent with REWSB  and neutralino LSP.  {\it Green} points form a subset of the {\it gray} {points}
and satisfy the sparticle and Higgs mass bounds, {as well as} all other constraints described in Section \ref{pheno}.}
\label{fig-5}
\end{figure}

Figure \ref{fig-5} shows the results in the $R_{tb\tau} - M_{1/2}$,  $R_{tb\tau} - m_{16_{1,2}}$, $R_{tb\tau} - m_{16_{3}}$ and
$m_{16_{3}} - \mu$  planes.
{\it Gray} points are consistent with REWSB  and neutralino LSP.  {\it Green} points form a subset of the {\it gray} {points}
and satisfy {the} sparticle and Higgs mass bounds along with all other constraints described in Section \ref{pheno}.

The $R_{tb\tau} - M_{1/2}$  plane shows the same interval for the parameter $M_{1/2}$ which is compatible with $t$-$b$-$\tau$ Yukawa unification as {previously} found with universal SSB sfermion {masses} \cite{Baer:2012cp}. This result was expected since {the} different SSB mass terms for the {first/second and the third families} do not significantly affect the RGE running and threshold corrections to the third generation fermions which is very crucial for $t$-$b$-$\tau$ Yukawa unification. {In this scenario, we do not find {acceptable} solutions with LSP neutralino as the correct dark mater candidate.}

\begin{figure}[]
\centering
{\label{fig:6}{\includegraphics[scale=0.4]{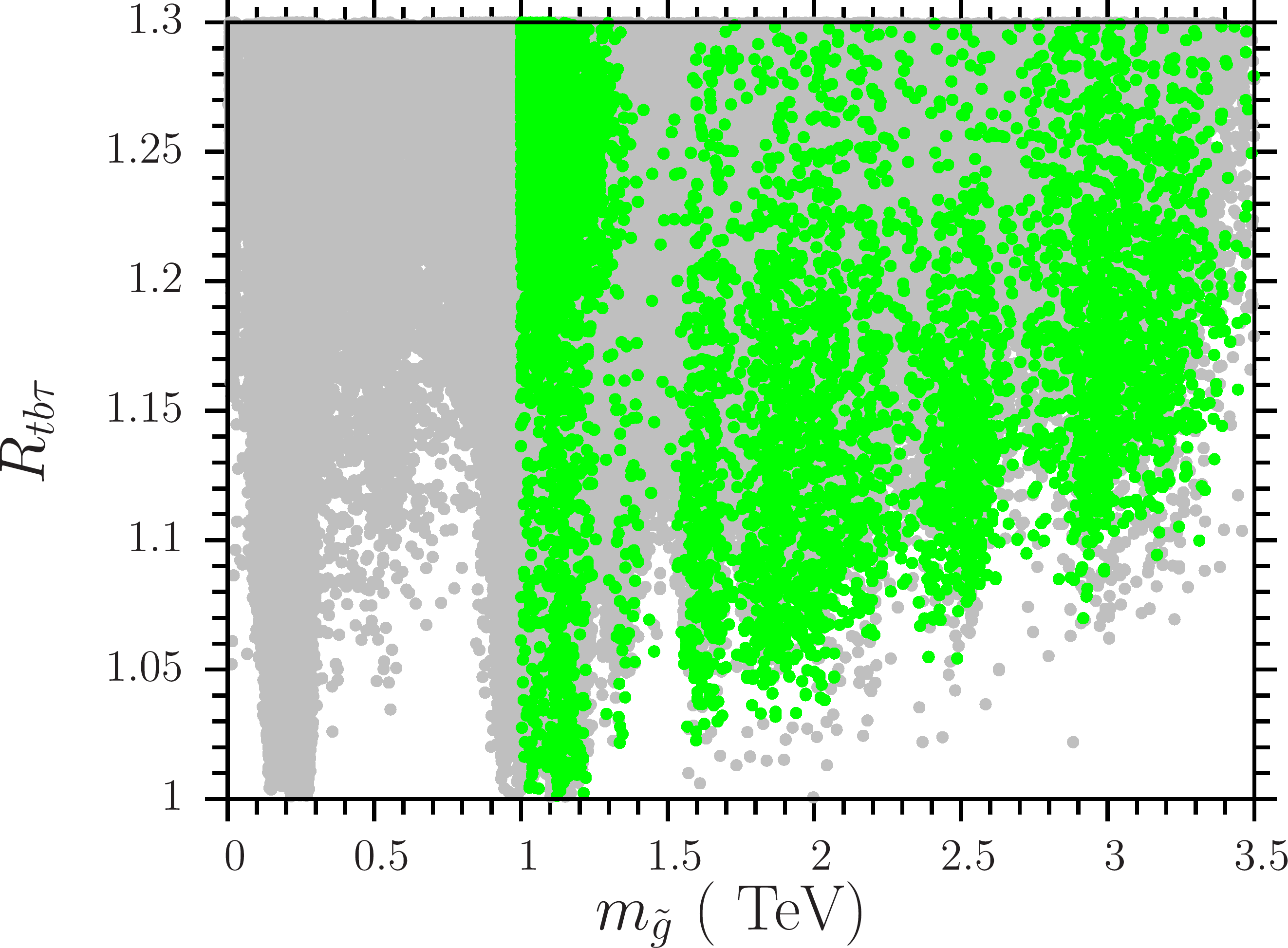}}}
{\label{fig:6}{\includegraphics[scale=0.4]{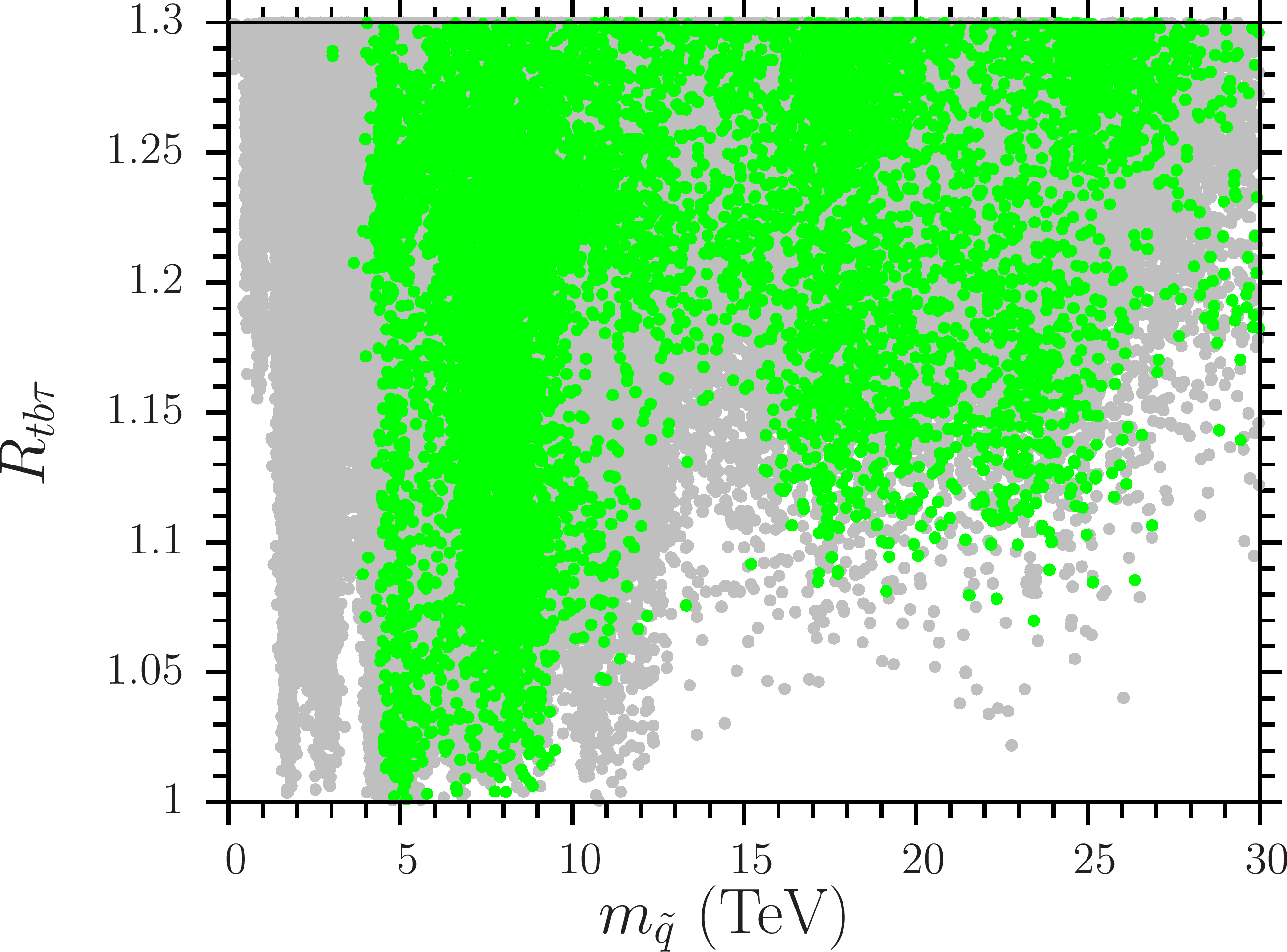}}}
\caption{ Plots in the $R_{tb\tau} - m_{\tilde g}$ and  $R_{tb\tau} - m_{\tilde q}$   planes. Color coding same as in Figure 5.}
\label{fig-6}
\end{figure}

\begin{table}[]\vspace{1.5cm}
\centering
\begin{tabular}{|p{3cm}|p{3cm}p{3cm}p{3cm}|}
\hline
\hline
                 	&	 Point 1 	&	 Point 2 	&	 Point 3 	\\
\hline							

$m_{16_{1,2}}$         	&$	7037.9	$&$	8195.9	$&$	22593.1	$\\
$m_{16_{3}}$         	&$	22196.8	$&$	25916.7	$&$	24856.1	$\\
$M_{1/2} $         	&$	465.6	$&$	644.4	$&$	1046.3	$\\
$A_0/m_{16_{3}}$         	&$	-2.2	$&$	-2.1	$&$	-2.1	$\\
$\tan\beta$      	&$	51.0	$&$	51.1	$&$	51.8	$\\
$m_{H_d}$         	&$	29104.8	$&$	34726.1	$&$	33575.9	$\\
$m_{H_u}$         	&$	25514.3	$&$	30738.2	$&$	29095.7	$\\
		  		  		  	
\hline		  		  		  	
$\mu$            	&$	7656	$&$	4788	$&$	5522	$\\

\hline		  		  		  	
$m_h$            	&$	{\bf 125.8}	$&$	{\bf 125.9}	$&$	{\bf 124.6}	$\\
$m_H$            	&$	3386	$&$	3700	$&$	8802	$\\
$m_A$            	&$	3364	$&$	3676	$&$	8745	$\\
$m_{H^{\pm}}$    	&$	3388	$&$	3701	$&$	8803	$\\
		  		  		  	
\hline		  		  		  	
$m_{\tilde{\chi}^0_{1,2}}$	&$	         281,          583	$&$	         372,          760	$&$	         562,         1119	 $\\

$m_{\tilde{\chi}^0_{3,4}}$	&$	        7522,         7522	$&$	        4724,         4724	$&$	        5461,         5462	 $\\

$m_{\tilde{\chi}^{\pm}_{1,2}}$	&$	         585,         7492	$&$	         762,         4683	$&$	        1123,         5421	$\\

$m_{\tilde{g}}$  	&$	{\bf 1567}	$&$	{\bf 2021}	$&$	{\bf 3005}	$\\
		  		  		  	
\hline $m_{ \tilde{u}_{L,R}}$	&$	        7025,         6507	$&$	        8199,         7617	$&$	       22665,        22481	$\\
                 		  		  		  	
$m_{\tilde{t}_{1,2}}$	&$	        3805,         6468	$&$	        4512,         7579	$&$	        4393,         7780	 $\\
                 		  		  		  	
\hline $m_{ \tilde{d}_{L,R}}$	&$	        7025,         7216	$&$	        8200,         8421	$&$	       22665,        22788	$\\
                 		  		  		  	
$m_{\tilde{b}_{1,2}}$	&$	        6546,         7370	$&$	        7699,         8752	$&$	        8037,         9480	 $\\
                 		  		  		  	
\hline		  		  		  	
$m_{\tilde{\nu}_{1}}$	&$	6637	$&$	7736	$&$	22451	$\\
                 		  		  		  	
$m_{\tilde{\nu}_{3}}$	&$	16532	$&$	19314	$&$	18357	$\\
                 		  		  		  	
\hline		  		  		  	
$m_{ \tilde{e}_{L,R}}$	&$	        6634,         7629	$&$	        7733,         8868	$&$	       22441,        22866	 $\\
                		  		  		  	
$m_{\tilde{\tau}_{1,2}}$	&$	       16487,         8172	$&$	       19261,         9555	$&$	       18335,         9020	 $\\
                		  		  		  	
\hline		  		  		  	
$\Delta a_{\mu}$  	&$	  1.45\times 10^{-11}	$&$	  1.82\times 10^{-11}	$&$	  2.88\times 10^{-12}	$\\

$\sigma_{SI}({\rm pb})$	&$	  2.31\times 10^{-14}	$&$	  6.16\times 10^{-15}	$&$	  1.84\times 10^{-15}	$\\

$\sigma_{SD}({\rm pb})$	&$	  2.53\times 10^{-12}	$&$	  1.10\times 10^{-10}	$&$	  1.39\times 10^{-10}	$\\

                		  		  		  	
\hline		  		  		  	
		  		  		  	
$R_{t b \tau}$     	&$	1.03	$&$	1.04	$&$	1.11	$\\

\hline
\hline
\end{tabular}
\caption{Benchmark points with good Yukawa unification and $m_h \sim$ 125 GeV. The points are shown with increasing gluino mass from point 1 to 3.} 
\label{table:2}
\end{table}

The  $R_{tb\tau} - m_{16_{1,2}}$ plane shows how low the SSB mass term for the first and second generation sfermions ($m_{16_{1,2}}$) can {become if} they are independent from  $m_{16_{3}}$.  We can compare these observations with the $R_{tb\tau} - m_{16_{3}}$ plane and note that the parameter $m_{16_{1,2}}$  can be 4-5 times lighter than $m_{16_{3}}$. However, $m_{16_{1,2}}$ lighter than 4 TeV is difficult if {the} various {experimental} constraints are implemented.

 The large difference between $M_{1/2}$ and $m_{16_{1,2}}$ (or $m_{16_{3}}$)  values, if we require $10\%$ or better unification, shows that the neutralino coannihilation scenario is not possible in order to yield the correct neutralino dark matter relic abundance.
Since $m_{16_{1,2}} \gtrsim 4$ TeV, there can be no significant contribution to the muon $g-2$ anomalous magnetic moment like we had in the previous section.
We also learn from {the} $m_{16_{3}} - \mu$ panel that the $\mu$-term is {greater} than 3 TeV if we demand $10\%$ or better Yukawa unification, ({\it green} points). {Comparing} this result with the gaugino mass interval obtained from demanding  $10\%$ or better unification, we conclude that the bino-higgsino mixed dark matter {scenario is not viable here}.

In Figure \ref{fig-6} we display  the results in the $R_{tb\tau} - m_{\tilde g}$ and  $R_{tb\tau} - m_{\tilde q}$  planes, {with} color coding  the same as in Figure \ref{fig-5}.
The $R_{tb\tau} - m_{\tilde g}$ panel shows that $t$-$b$-$\tau$ Yukawa unification predicts an upper bound on the gluino mass which can {easily} be tested at LHC14.  The result in green color from the $R_{tb\tau} - m_{\tilde q}$ plane shows that in this model squarks will be difficult to {find} at LHC14, but there is some hope that {they} might  be accessible at LHC33 \cite{cms_lim}.

In Table \ref{table:2} we show three benchmark points for this scenario {which display} good Yukawa unification with {the required} Higgs mass. In addition all other constraints described in Section \ref{pheno} are satisfied. As {previously mentioned}, this SO(10) model does not exhibit coannihilation and the contribution to the $g-2$ anomaly is also not significant.  The gluino is the lightest colored sparticle for the three points and may be found at the LHC.

\section{ Conclusion \label{conclusions}}

We {discussed} {supersymmetric} SO(10) grand {unification} with { {non-universal}} and { universal} gaugino masses at $M_{\rm GUT}$ {with the sfermion masses of the first and second generations} different from that of the third generation. We explored {the consistency of} good $t$-$b$-$\tau$ Yukawa unification in these models with various experimental observations, namely, the Higgs and sparticle mass limits, B-physics constraints, WMAP relic density bound and the muon anomalous magnetic moment. We further studied the sparticle spectroscopy of these models and listed some benchmark scenarios that can be explored at 14 TeV LHC.

{In the scenario with non-universal gaugino masses, the soft supersymmetry breaking parameters $M_i$ ($i$=1, 2, 3) are treated as independent}. In this case all of the above mentioned constraints can be satisfied. The colored sparticles {are all} found to be very heavy ($\gtrsim 5$ TeV) for 10\% or better Yukawa unification. The sleptons (smuon and stau) in this case can be as light as 200 GeV. The correct relic abundance for neutralino dark matter is realized through various channels including {neutralino-stau(smuon)} coannihilation and $A$ resonance.

The second model {has} universal gaugino masses and non-universal Higgs masses at $M_{\rm GUT}$. The gluino {turns out to be} the lightest colored sparticle {with mass} $\gtrsim$ 1.5 TeV. The sfermions including the sleptons, {however}, {are all} very heavy ($\gtrsim 4$ TeV), {so that the muon $g-2$ anomaly is unresolved}. The {LSP neutralino}
{in this case is not a viable dark matter candidate}. 
 {The remaining experimental} constraints are satisfied in this scenario, and {we present some benchmark points}. {They} exhibit {acceptable} Yukawa unification and the gluino is the only sparticle accessible at the LHC.


\section*{Acknowledgments}

This work is supported in part by the DOE Grant No. DE-FG02-12ER41808.  This work used the Extreme Science
and Engineering Discovery Environment (XSEDE), which is supported by the National Science
Foundation grant number OCI-1053575. I.G. acknowledges support from the  Rustaveli
National Science Foundation  No. 03/79.


\end{document}